\documentclass[smallabstract,smallcaptions]{dccpaper}

\usepackage{epsfig}
\usepackage{citesort}
\usepackage{amsmath}
\usepackage{amssymb}
\usepackage{color}
\usepackage{url}

\usepackage{cite}
\usepackage{verbatim}
\usepackage{algorithmic}
\usepackage{textcomp}
\usepackage{xcolor}
\usepackage{multirow}
\usepackage{multicol}
\usepackage{makecell}
\usepackage{graphicx} 
\usepackage{float} 
\usepackage{subfigure} 
\usepackage[breaklinks=true,bookmarks=false]{hyperref}
\usepackage{booktabs}
\usepackage[ruled,vlined]{algorithm2e}
\usepackage{titlesec} 
\titlespacing{\section}{0pt}{2.5ex plus 1ex minus .2ex}{1.3ex plus .2ex}
\titlespacing{\subsection}{0pt}{2.5ex plus 1ex minus .2ex}{1.3ex plus .2ex}
\newlength{\figurewidth}
\newlength{\smallfigurewidth}

\begin{document}
\begin{center}
\large
\textbf{Video Compression with Arbitrary Rescaling Network}

\end{center}

\begin{center} 
Mengxi Guo, Shijie Zhao\footnotemark{*}, Hao Jiang, Junlin Li and Li Zhang\\
{\small\begin{minipage}{\linewidth}\begin{center}
\begin{tabular}{c}
Bytedance Inc. \\
San Diego, CA, 92122 USA and Shenzhen, China \\
$\left.\{guomengxi.qoelab,zhaoshijie.0526,jianghao, lijunlin.li,lizhang.idm \right.\}@bytedance.com$
\end{tabular}
\end{center}\end{minipage}}
\end{center}
\begin{center}
\end{center}
\begin{abstract}
Most video platforms provide video streaming services with different qualities, and the quality of the services is usually adjusted by the resolution of the videos. So high-resolution videos need to be downsampled for compression. In order to solve the problem of video coding at different resolutions, we propose a rate-guided arbitrary rescaling network (RARN) for video resizing before encoding. To help the RARN compatible with standard codecs and generate compression-friendly results, an iteratively optimized transformer-based virtual codec (TVC) is introduced to simulate the key components of video encoding and perform bitrate estimation. By iteratively training the TVC and the RARN, we achieved 5\%-29\% BD-Rate reduction anchored by linear interpolation under different encoding configurations and resolutions, exceeding the previous methods on most test videos. Furthermore, the lightweight RARN structure can process FHD (1080p) content at real-time speed (91 FPS) and obtain a considerable rate reduction.
\end{abstract}

\footnotetext[1]{Corresponding author.}
\section{Introduction}
The transmission of high-resolution videos consumes a lot of backbone network bandwidth and puts massive pressure on the redistribution points. Streaming media technology providers industriously launch Dynamic Adaptive Streaming over HTTP (DASH) to alleviate network pressure, which means the server has to convert the resolution of videos to generate encoded files of different bitrates. The widely used sampling method, like bicubic interpolation, is efficient but not designed for video coding, and key information might be lost during sampling. 


With the development of machine learning, some downsampling compression schemes \cite{ho2021rr, son2021enhanced, wei2021video} apply neural networks at pre- and post-processing to improve the rate-distortion performance of traditional codecs. However, the application of these algorithms faces three major problems. First, many contents on the recent popular media platform are created by users, which vary widely in size. To scale these irregular-sized videos to a standard resolution, it is essential to use an arbitrary sampling method, and most algorithms only work under a fixed configuration. Second, most of those schemes need a super-resolution network as post-processing to improve the quality of the decoded videos, introducing tremendous computations and seriously affecting the users' experience. Last, during the training phase of the down-sampling network, previous precoding methods do not thoroughly consider the quality degradation of video frames raised by the encoding and decoding process.

To address these problems, we propose a practical downsampling compression scheme to use a neural network as a pre-processing module for traditional codecs to improve compression performance. It is compatible with standard codecs when deploying. Specifically, we propose a rate-guided arbitrary rescaling network (RARN) to achieve video resampling to alleviate the overhead of model deployment and complex bandwidth scenarios. During the resampling, we use a pre-trained variational autoencoder (VAE) module adopted in image compression task \cite{balle2018variational} for rate estimation. The bitrate information is delivered to the self-attention layer to guide the sampling process. To make RARN aware of the compression distortion caused by the encoding and decoding process, we propose a transformer-based virtual codec (TVC) to simulate the RD performance of standard codecs. The key component of TVC is a swin-transformer-based invertible neural network (swin-INN). The swin-INN allows the TVC to learn the distortion from standard codecs, and the cyclic shift attention \cite{liu2021swin} in the network can approximate the prediction modes of HEVC \cite{sullivan2012overview}. 
We evaluate our method with the standard codecs HEVC and VVC following linear upsampling post-processing filters. Experimental results show that our proposed method can achieve top performance for arbitrary sampling ratios in quantity and quality. 

The main contributions of our proposed method can be summarized as follows:
\begin{itemize}
\item Unlike previous precoding networks, which can only work under a fixed scaling ratio, we propose a rate-guided arbitrary rescaling network (RARN) that can resample video to arbitrary resolution while retaining key information in frames.


\item We propose a novel transformer-based virtual codec (TVC) to simulate the degradation process of encoding and decoding. The RARN can be trained with the TVC iteratively, which helps the precoding network to be aware of the compression and achieve better RD performance with standard codecs.

\item In the downsampling-based compression framework, our proposed RARN achieves better performance than previous methods in most test sequences. Besides, the lightweight model can achieve real-time inference speed (91 FPS) while maintaining satisfactory compression performance.
\end{itemize}

\section{Proposed Method}

\subsection{Problem Formulation}
\begin{figure}[t]
\begin{center}
\includegraphics[width=0.7\linewidth]{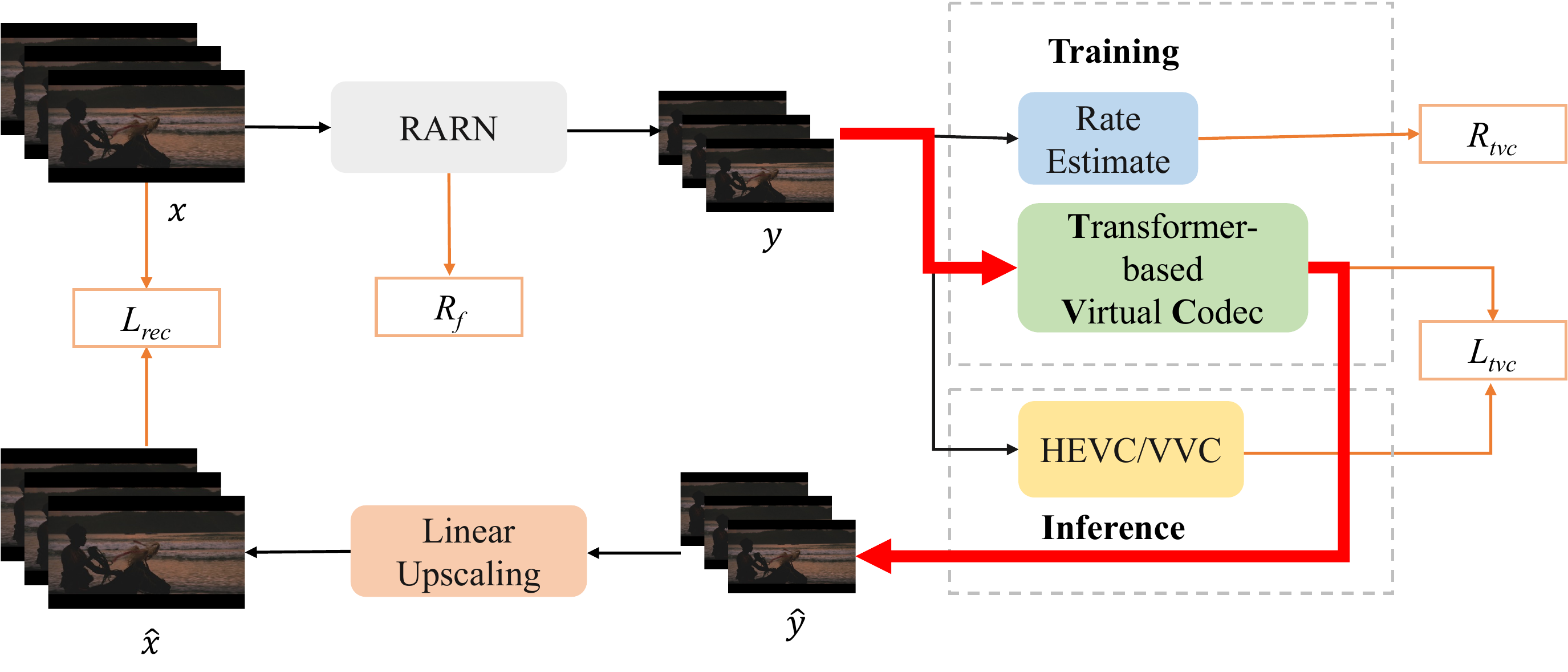}
\end{center}
\vspace{-0.8cm}
\caption{\centering  The architecture of the proposed framework. During training, the downsampled frames $y$ are fed into the Transformer base Virtual Codec along the red arrow direction.}
\label{overallarch}
\end{figure}

The overall architecture of the proposed downsampling-based video compression framework is shown in Fig.~\ref{overallarch}.
The original video $\textit{x}$ is processed into the low-resolution video $\textit{y}$ through the rate-guided arbitrary rescaling network (RARN), which assists in preserving essential structures and valid information:
\begin{equation}
    y=f(\mathbf{x}; \theta_{f})
    R_{f} = \Phi_{f}(x),
\end{equation}
where the $\textit{f}$ and $\theta_{f}$ represent the transformation function and trainable parameter respectively, and $R_{f}$ is the evaluated auxiliary bitrates information of raw frames.

In the training pipeline, we use a transformer-based virtual codec (TVC) to process low-resolution video into $\hat{y}$ to simulate the process of degradation during actual compression while estimating the bitrate of the frames:
\begin{equation}
    \hat{y}=g(\mathbf{y}; \theta_{g}),
    R_{TVC} = \Phi_{b}(y)
\end{equation}
where the $\textit{g}$ and $\theta_{g}$ represent the transformation function and trainable parameters of TVC, while the $R_{tvc}$ is the estimated bitrate and $\Phi$ is the rate estimate network.
We adopt a differentiable neural network as the TVC to provide accurate gradient estimation for back-propagation and permit the RARN to perceive the rate-loss information of actual encoding.

In the inference pipeline, the low-resolution video is encoded and decoded by the standard codecs, so the proposed RARN can be deployed completely independent of actual codecs and accomplish the goal of video quality enhancement and bitrate saving.
Finally, $\hat{y}$ is upsampled by a linear upscaling filter(e.g. Bicubic) to obtain the original resolution video $\hat{x}$.

The optimization goal of RARN is to use the least bitrates for the best video quality, which can be performed by minimizing the $R + \lambda D$ loss:
\begin{equation}
\begin{aligned}
 \boldsymbol{\theta}_{\boldsymbol{f}}^{*} \approx \underset{f, \Phi}{\operatorname{argmin}} {\Phi}_{b}(y) +\lambda \delta (x, \hat{x}),
\label{rdloss}
\end{aligned}
\end{equation}
 where the $\Phi_{b}(y)$ denotes the estimated bitrate of low-resolution frames $y$ and $\lambda$ is the quality loss function, and $\lambda$ controls the trade-off between rate and distortion.

\subsection{Arbitrary Resolution Transform Network}
\begin{figure}[htbp]
\centering
\subfigure[overall architecture]{
\begin{minipage}[t]{0.3\linewidth}
\centering
\includegraphics[width=2in]{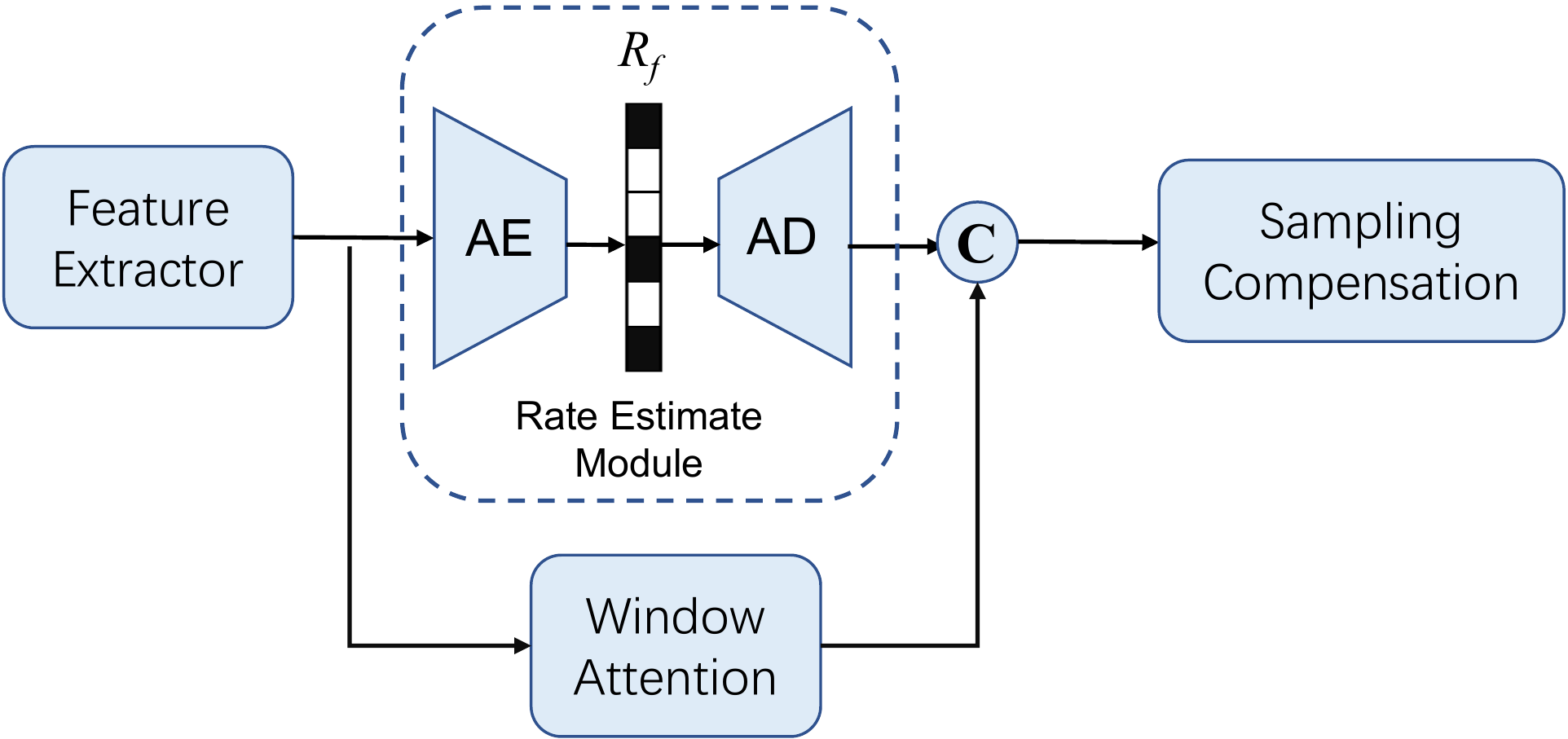}
\label{RARN}
\end{minipage}%
}
\subfigure[sampling compensation module]{
\begin{minipage}[t]{0.6\linewidth}
\centering
\includegraphics[width=3in]{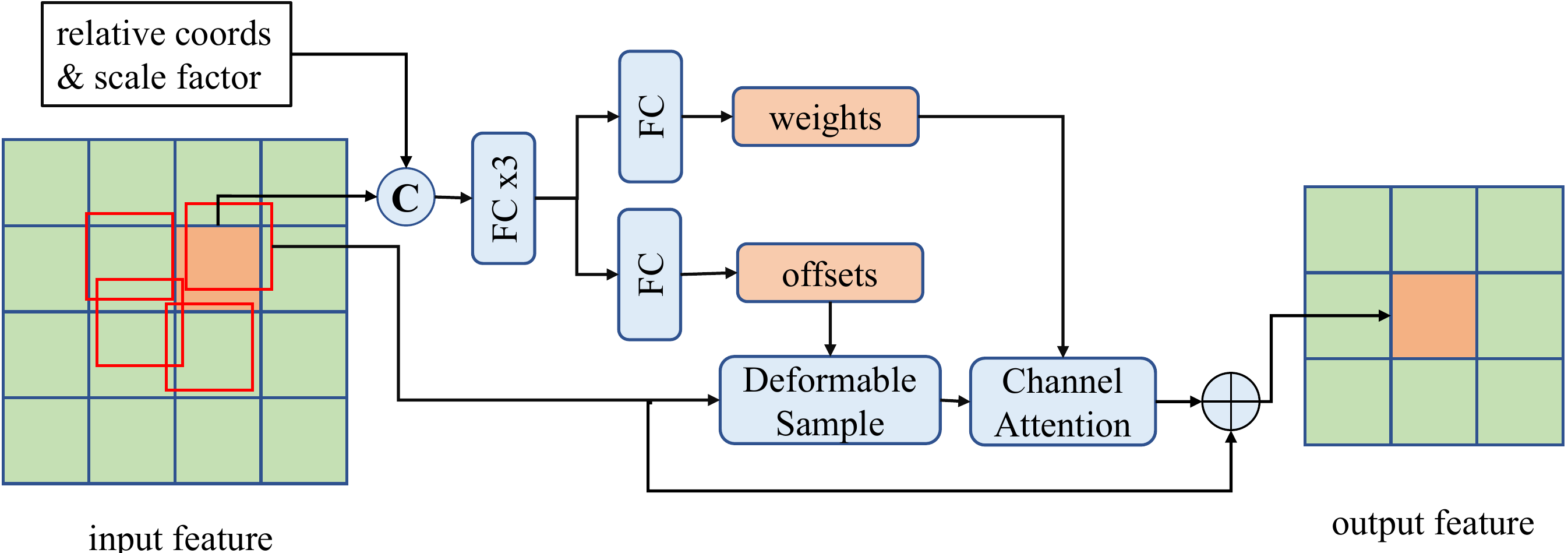}
\label{compensation}
\end{minipage}%
}
\caption{Rate-guided arbitrary rescaling network (RARN)}
\label{RARN-arch}
\end{figure}

We propose a rate-guided arbitrary rescaling network (RARN) as a precoding module in Fig.~\ref{RARN}, which works to reduce resolution while preserving fundamental structures and critical textures.
We transform the raw data into feature space using resblocks to alleviate the concern of pixel misalignment when sampling directly in pixel space.
Then a pre-trained VAE structure is adopted for rate estimation. 
The final output layer of the VAE is recast into features of multiple channels to characterize the bitrate information. There are two main purposes why we introduce the rate estimate module into the sampling process: (i) the spatial and temporal complexity can be introduced from the rate estimation, which can control the spatial sampling intensity; (ii) compared with \cite{wei2021video} which estimates the bitrates at the end of the pipeline, the evaluated auxiliary bitrates information of raw frames $R_{f}$ can directly guide the sampling module and void gradient vanishing.
Finally, we use a sampling compensation module with the estimated bitrate as prior information to adaptively sample features processed by self-attention transformation.

The structure of the sampling compensation module is shown in Fig.~\ref{compensation}. We use grid\_sample combined with deformable offset for resolution conversion and compensation.
We use the normalized coordinate to calculate the corresponding input coordinate and query the input features, where we adopt the bicubic interpolation as implementation.
Assuming that the output integer coordinates are $(R, C)$, where $R \in \left[0, H-1\right]$ and $C \in \left[0, W-1\right]$, we normalize the coordinates to $(r, c)$, where $r, c \in \left[-1, 1\right]$ using the coordinate normalize function $p$:
\begin{align}
      r &= p(R, H) = -1.0 + \frac{2R + 1}{H} ,
\end{align}
taking the row dimension of coordinates as an example.
The normalized output coordinates are projected to the input integer coordinates $\left[\hat{R}, \hat{C}\right]$ by multiplying the input size $\left[\hat{H}, \hat{W}\right]$ and query the input features $F^{query}_{\hat{R}, \hat{C}}$.
We can calculate a normalized coordinate sampling error $\left[E(R), E(C)\right]$ according to the scale factor $S$, which supplies supplementary information for sampling compensation:
\begin{align}
    E(R) = r - p(round(\frac{r}{S} * \hat{H}), H), \quad r = p(R, H).
\end{align}
The queried input feature, normalized coordinate sampling error, and scale factor are concatenated together and fed into a double-headed MLP, which predicts deformable offset $\left[\delta_{R}, \delta_{C}\right]$ and channel compensation weights $W$.
The deformable $grid\_sample$ operation resamples the demanded input features for each output coordinate and the channel-attention mechanism adds the sampling compensation to the initial queried features to obtain the actual output:
\begin{equation}
    F^{out}_{R, C} = F^{query}_{\hat{R}, \hat{C}} + \sum_{k=1}^{K} w_k \cdot F_{\hat{R} + \delta_{Rk}, C + \delta_{Ck}},
\end{equation}
where $K$ represents the number of deformable offsets and $w_{k}$ is the channel compensation weights.

\begin{figure}[t]
\begin{center}
\includegraphics[width=0.9\linewidth]{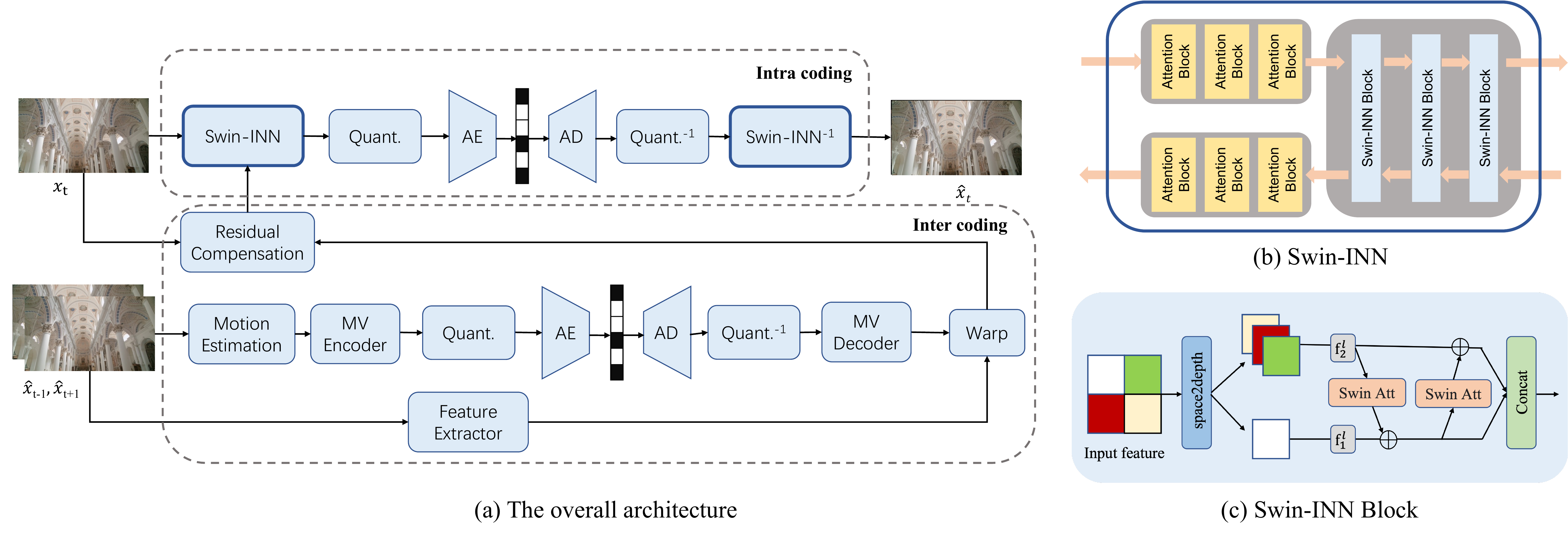}
\end{center}
\vspace{-0.8cm}
\caption{The architecture of transformer-based virtual codec.}
\label{swinTVC}
\end{figure}

\subsection{Transformer-based Virtual Codec}
We use a transformer-based virtual codec (TVC) in Fig.~\ref{swinTVC}(a) instead of standard codecs when training the RARN, enabling the gradients to be propagated back to the RARN.
TVC is iteratively updated according to the output of the standard codec to simulate the behaviour of video encoding and provide accurate gradient estimation for joint training.
In actual encoding scenarios, the image type of B frame counts the highest proportion, so we adopt the structure of $GOP=3$ to make up the structure. Thus the TVC can simulate the basic implementation of inter-frame and intra-frame coding with motion estimation and motion compensation sufficiently.

For intra-frame coding, we use a swin-transformer-based invertible neural network (swin-INN) to model the intra-coding transformation.
As shown in Fig.~\ref{swinTVC}(b), our swin-INN consists of three attention blocks for feature enhancement and stacked cyclic shift attention modules \cite{liu2021swin} so that each window can combine data on the left and above to approximate the intra-prediction of HEVC when calculating self-attention.
We use a smaller window size $k=8$, the smallest coding unit size in HEVC, to reduce the computation burden and facilitate more flexibility in assigning different bitrates to different parts of the image.
The reversible structure (Fig.~\ref{swinTVC}(c)) of INNs contributes to a tractable Jacobian mathematically, which makes it simpler to model data distributions by computing probability integrals but limits its nonlinear transformation capacity \cite{dinh2014nice}. 
Therefore, we add extra non-local attention blocks before the swin-INN blocks to enhance the nonlinear transformation capability. 

For inter-frame coding, we use the structure of picture type B frame, which restores the current frame $x_{t}$ based on the two adjacent frames $x_{t-1}, {x}_{t+1}$.
We perform motion estimation and motion compensation (MEMC) in the feature domain to mitigate the impact of inaccurate optical flow estimation.
We use RAFT \cite{teed2020raft} model for motion estimation and perform motion compensation to obtain the current predicted frame:
\begin{equation}
f_{\text {predict }}\left(\hat{x}_{t-1, t+1}\right)=\operatorname{warp}\left(f_{fea}\left(\hat{x}_{t-1, t+1}\right), \hat{m}_t\right),
\end{equation}
where $fea$ represents the feature extractor and $\hat{m}_t$ represents the motion vectors.
We use conditional convolution to automatically explore the correlation between the predicted frame and the current frame to eliminate redundancy instead of a simple subtraction operation:
\begin{equation}
\hat{x}_t=f_{\text {swin }}\left(f_{\text {residual }}\left(x_{t} \mid \widetilde{x}_t\right) \right) 
\text { with } \widetilde{x}_t=f_{\text {predict }}\left(\hat{x_{t-1}}\right) ,
\end{equation}
$f_{\text {swin }}$ represents the coding process, and $f_{\text {residual }}$ is the conditional residual extractor.
This method can exert the ability of content adaptation of the network, especially when the current frame cannot find a good reference in previously decoded frames.

\subsection{Training Strategy}
The main purpose of the proposed video precoding algorithm is to obtain high video quality at a low bitrate, so the training loss mainly includes two indicators:
\begin{equation}
    L = D + \lambda R,
\end{equation}
$\lambda$ controls the trade-off between distortion metrics D and bitrate loss R.
Distortion is composed of the L2 loss between the recovered frame and the original frame and the MSE between the Lanczos down-sampled frame and network-processed frame to limit the output of the network and ensure that joint training is not corrupted:
\begin{equation}
\begin{aligned}
D &= L_{rec} + L_{f}
&= \left\|\mathbf{x}-\mathrm{Bicubic}(\mathbf{\widetilde{y}})\right\|_{2}^{2} + \lambda_{d} \left\|f(\mathrm{x})-\mathrm{Bicubic}(\mathrm{x})\right\|_{2}^{2},
\end{aligned}
\end{equation}
The rate loss is calculated by the rate estimation module in the TVC and the rate auxiliary information in the RARN:
\begin{equation}
\begin{aligned}
R &= R_{tvc} + \lambda_{r} R_{f}&= \Phi_{b}(y) + \lambda_{r} \Phi_{f}(x).
\end{aligned}
\end{equation}

Furthermore, we use the output of the actual codec $\hat{x}$ to iteratively perform the training of the TVC at each iteration:
\begin{equation}
\mathcal{L}_{tvc} = \left\|\mathrm{HEVC(x)}-\Phi(\mathrm{x})\right\|_{2}^{2}.
\end{equation}

\section{Experimental Results}

\begin{table}[t]
\begin{center}
\caption{\label{xiph_265}\centering Average BD-rate results in XIPH FHD test sequence}
\label{tab1}
{
\begin{tabular}{ccccc}
\hline 
{\makecell[c]{libx265, preset=medium \\ GOP=30, bicubic}} & \multicolumn{2}{c}{Bourtsoulatze \cite{bourtsoulatze2019deep}} 
& \multicolumn{2}{c}{$\mathrm{Proposed}$} \\
\cline { 2 - 5 } & Bicubic & Lanczos & Bicubic & Lanczos \\
\hline 
{\makecell[c]{scale factor = 5/2}} 
& $-25.17\%$ & $ -18.84\% $ & $\textbf{-29.27\%}$ & $\textbf{-20.15\%}$ \\
\hline
{\makecell[c]{scale factor = 2}} 
& $-19.25\%$ & $-14.46\%$ & $\textbf{-21.57\%}$ & $\textbf{-16.24\%}$\\
\hline
{\makecell[c]{scale factor = 3/2}}  
& $\textbf{-13.18\%}$ & $-8.26\%$ & $-11.87\%$ & $\textbf{-10.57\%}$ \\
\hline
\end{tabular}
}
\end{center}
\end{table}

\begin{table}[t]\small
\centering
\caption{ \centering 
BD-rate comparison of down-sampled based compression frameworks in HEVC all-intra configuration}
\label{HEVC}
\scalebox{0.9}{
\begin{tabular}
{cccccccc}
\hline \hline & & RR-DnCNN v2.0 \cite{ho2021rr} & Hanbin Son \cite{son2021enhanced} & Yuzhuo \cite{wei2021video} & Proposed\\
\hline
\multirow{2}{*}{ClassA}
&Traffic & $-15.2 \%$ & $-15.0 \%$ & $\textbf{-16.6 \%}$ & $-7.3 \%$ \\
& PeopleOnStreet & $-16.3 \%$ & $-15.4 \%$ & $\textbf{-16.1 \%}$ & $-6.3 \%$ \\

\hline \multirow{5}{*}{ClassB} 
& Kimono & $-11.3 \%$ & $-11.7 \%$       & $\textbf{-13.1 \%}$  & $-2.2 \%$  \\
& ParkScene & $-11.7 \%$ & $-11.6 \%$     & $\textbf{-12.7 \%}$  & $-8.7 \%$ \\
& Cactus & $-9.9 \%$ & $-14.9 \%$         & $\textbf{-18.9 \%}$ & $-16.5 \%$  \\
& BasketballDrive & $-0.4 \%$ & $-11.3 \%$ & $-12.4 \%$  & $\textbf{-31.2 \%}$ \\
& BQTerrace & $-$ & $-14.1 \%$ & $-$ & $\textbf{-28.2 \%}$ \\

\hline \multirow{4}{*}{ ClassC } 
& BasketballDrill & $-12.5 \%$ & $-19.2 \%$ & $-18.1 \%$ & $\textbf{-25.9 \%}$  \\
& BQMall & $-2.6 \%$ & $-7.3 \%$          & $-5.8 \%$  & $\textbf{-20.6 \%}$ \\
& PartyScene & $-0.9 \%$ & $-9.3 \%$       & $-8.4 \%$ & $\textbf{-23.0 \%}$ \\
& RaceHorcesC & $-8.5 \%$ & $-14.2 \%$    & $-13.6 \%$  & $\textbf{-22.4 \%}$ \\

\hline \multirow{4}{*}{ ClassD } 
& BasketballPass & $-$ & $\textbf{-20.1 \%}$ & $-$ & $-19.8 \%$ \\
& BQSquare & $-$ & $-15.4 \%$ & $-$ & $\textbf{-30.0 \%}$ \\
& BlowingBubbles & $-$ & $\textbf{-19.7 \%}$ & $-$ & $-18.8 \%$ \\
& RaceHorsesD & $-$ & $-18.5 \%$  & $-$  & $\textbf{-20.8 \%}$\\

\hline \multirow{3}{*}{ ClassE } 
& FourPeople & $-9.0 \%$ & $-15.2 \%$ & $-10.8 \%$ & $\textbf{-19.2 \%}$  \\
& Johnny & $-12.5 \%$ & $\textbf{-20.9 \%}$ & $-13.3 \%$ & $-19.0 \%$ \\
& KristenAndSara & $-3.8 \%$ & $-15.7 \%$ & $-8.0 \%$ & $\textbf{-27.4 \%}$ \\

\hline 
\multirow{5}{*}{Summary}
& Class A & $-15.7 \%$ & $-15.2 \%$  & $\textbf{-16.4 \%}$  & $-7.8 \%$\\
& Class B & $-8.52 \%$ & $-12.7 \%$  & $-14.1 \%$ & $\textbf{-17.4 \%}$  \\
& Class C & $-6.1 \%$ & $-12.5 \%$ & $-11.5 \%$ & $\textbf{-22.9 \%}$ \\
& Class D & $-$ & $-18.4 \%$  & $-$ & $\textbf{-22.4 \%}$\\
& Class E & $-5.9 \%$ & $-17.2 \%$ & $-10.7 \%$ & $\textbf{-21.9 \%}$ \\

\hline \multicolumn{2}{c}{Overall} & $-10.7 \%$ & $-15.2 \%$ & $-12.93 \%$ &$\textbf{-19.4 \%}$ \\
\hline \hline
\end{tabular}}
\end{table}

\begin{figure}[htbp]
\centering
\subfigure[BD-SSIM of XIPH]{
\begin{minipage}[t]{0.23\linewidth}
\centering
\includegraphics[width=1.3in]{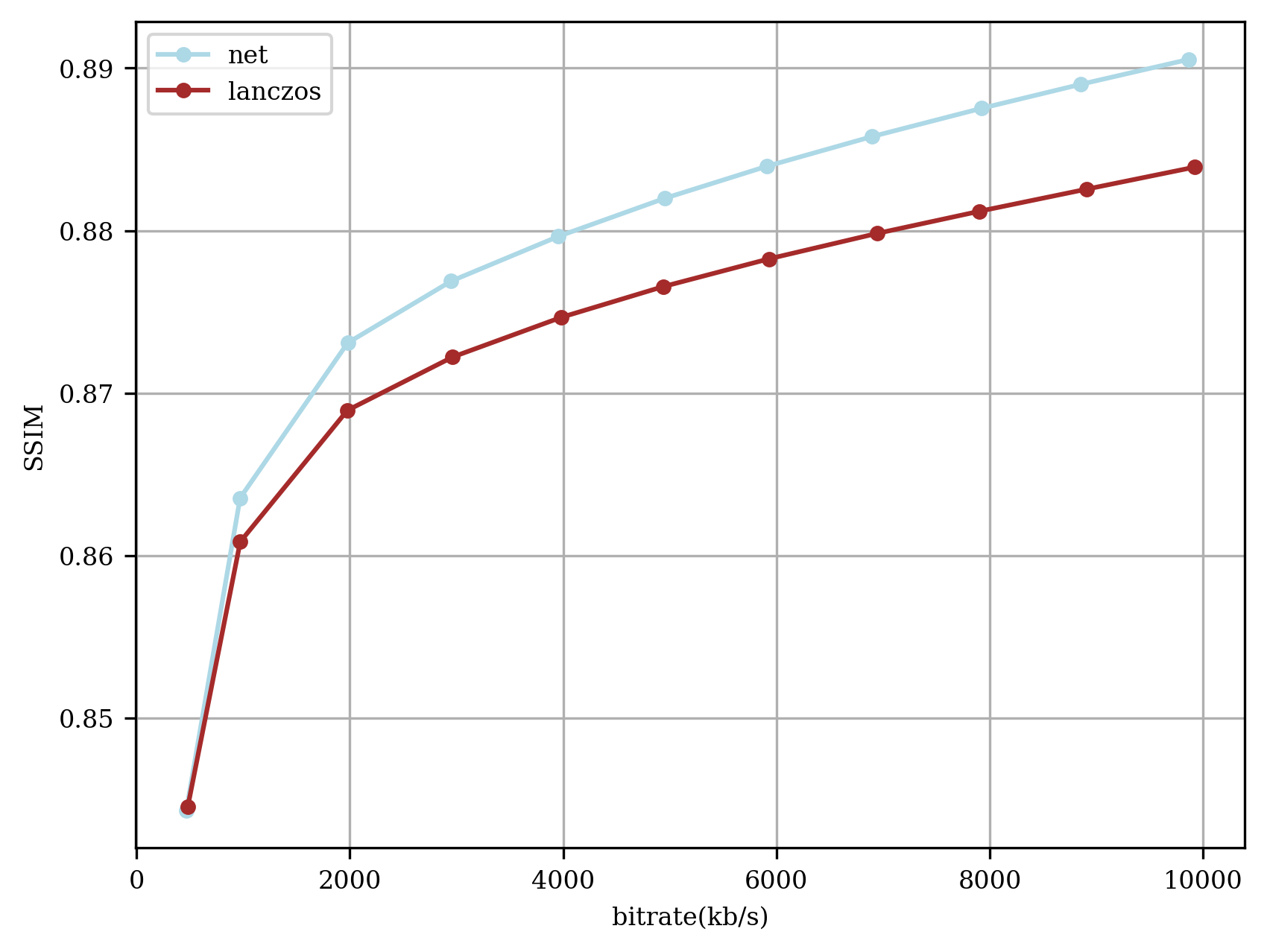}
\label{xiph-ssim}
\end{minipage}%
}
\subfigure[BD-PSNR of XIPH]{
\begin{minipage}[t]{0.23\linewidth}
\centering
\includegraphics[width=1.3in]{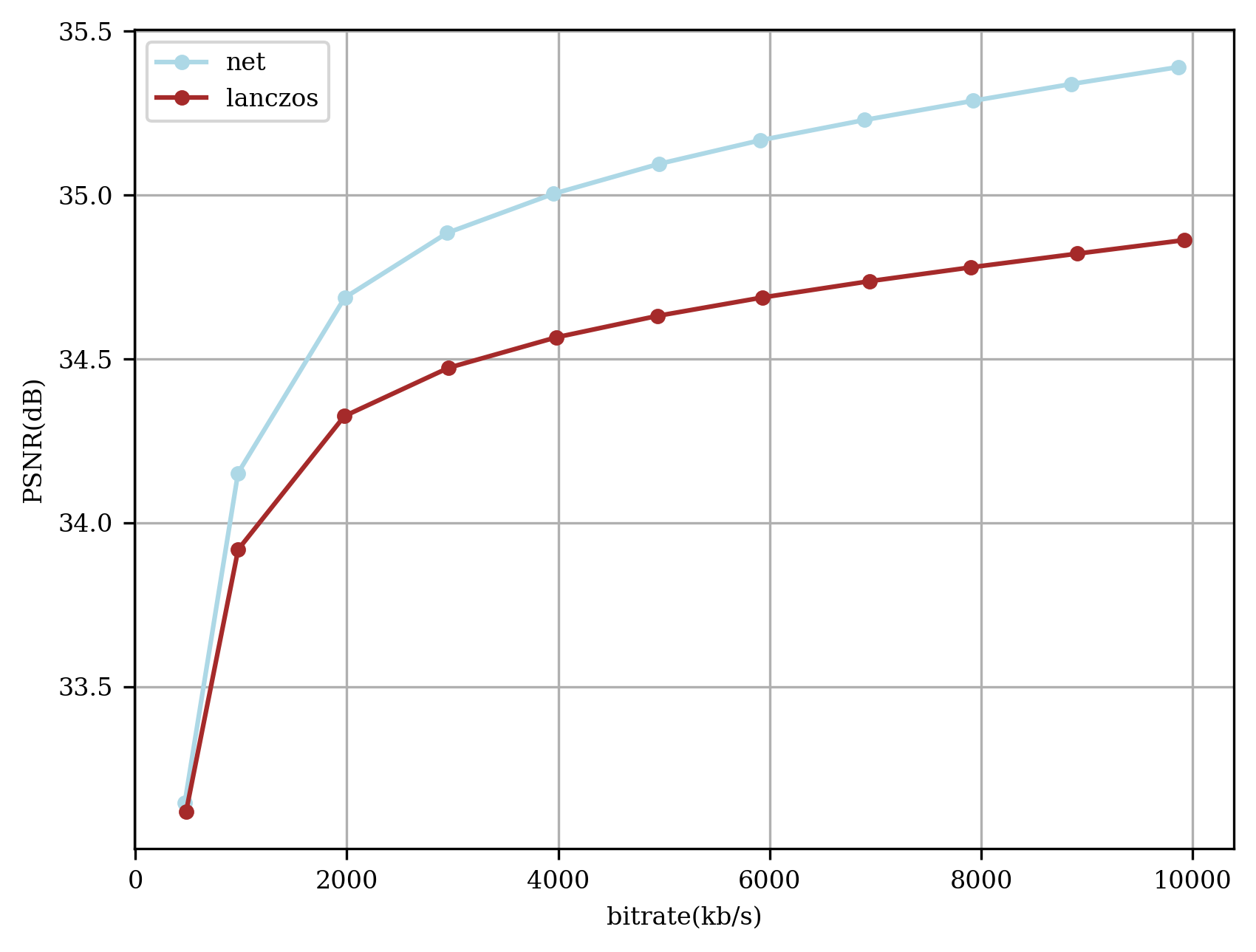}
\label{xiph-psnr}
\end{minipage}%
}
\subfigure[BD-SSIM of HEVC]{
\begin{minipage}[t]{0.23\linewidth}
\centering
\includegraphics[width=1.3in]{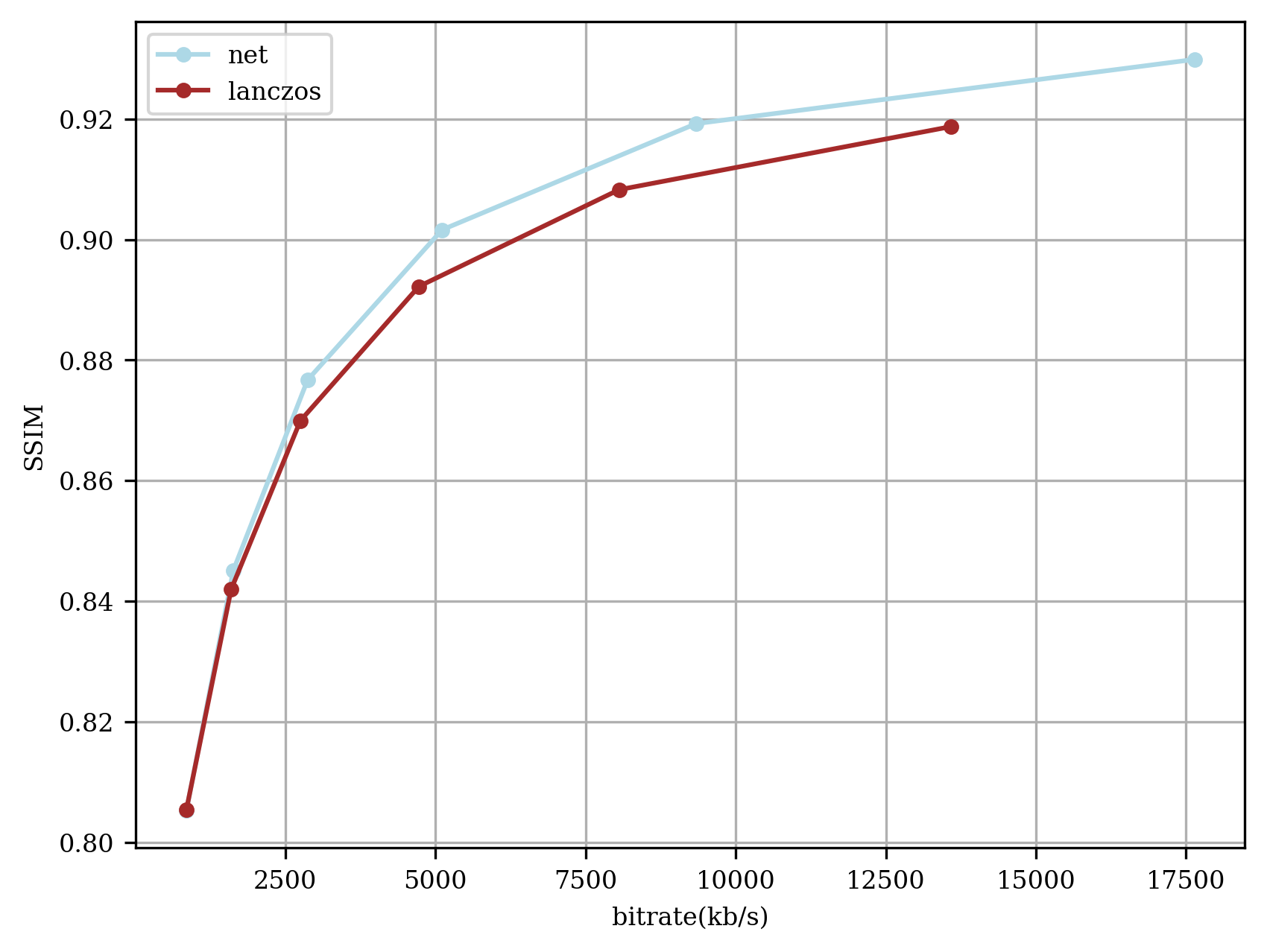}
\label{hevc-ssim}
\end{minipage}%
}
\subfigure[BD-PSNR of HEVC]{
\begin{minipage}[t]{0.23\linewidth}
\centering
\includegraphics[width=1.3in]{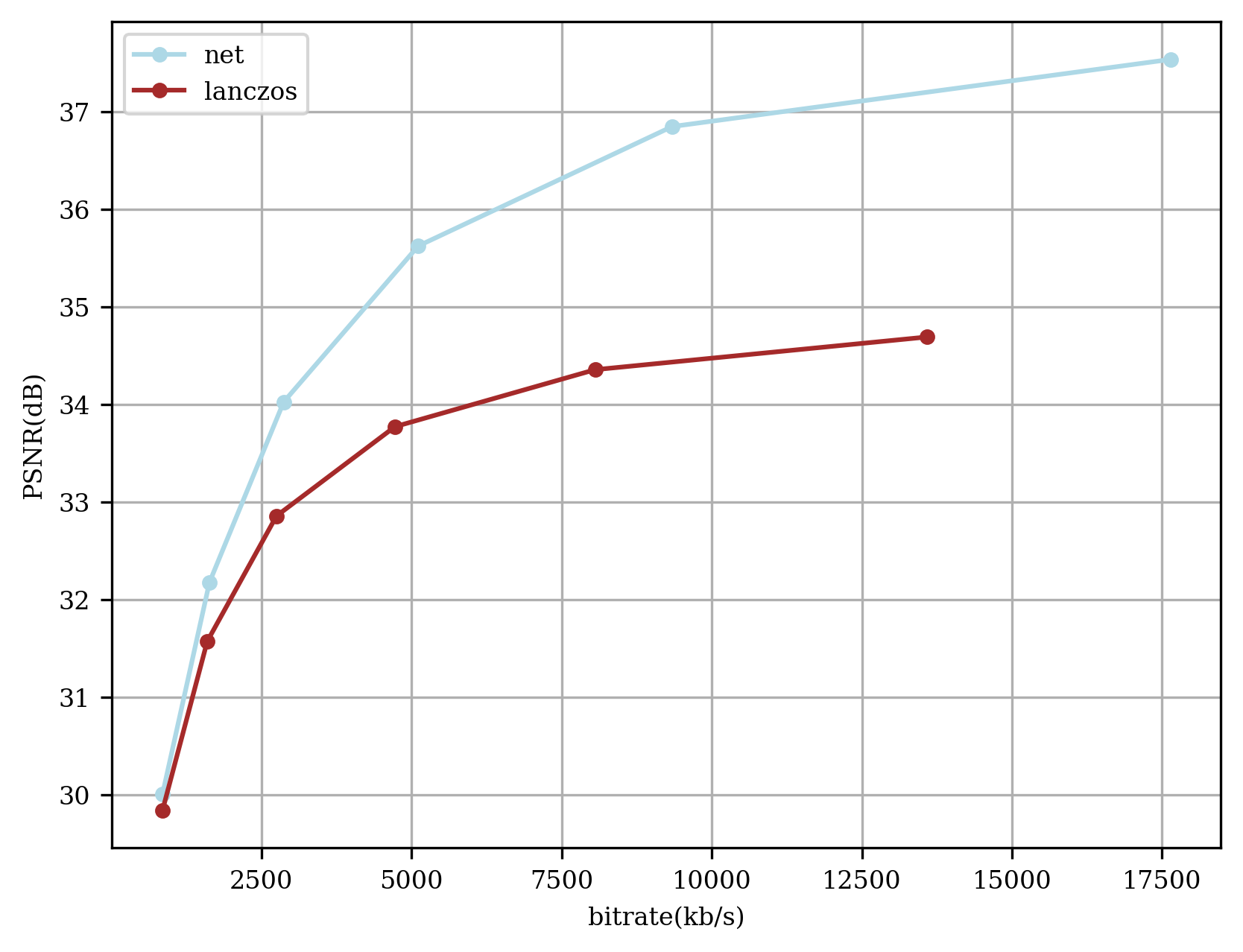}
\label{hevc-psnr}
\end{minipage}%
}
\caption{\centering  Comparison of proposed method and liner filter baseline on XIPH dataset and HEVC ClassB sequences of H.265/HEVC. }
\label{bdcurve}
\end{figure}

\begin{figure}[htbp]
\centering
\subfigure[BD-SSIM of XIPH]{
\begin{minipage}[t]{0.4\linewidth}
\centering
\includegraphics[width=1.3in]{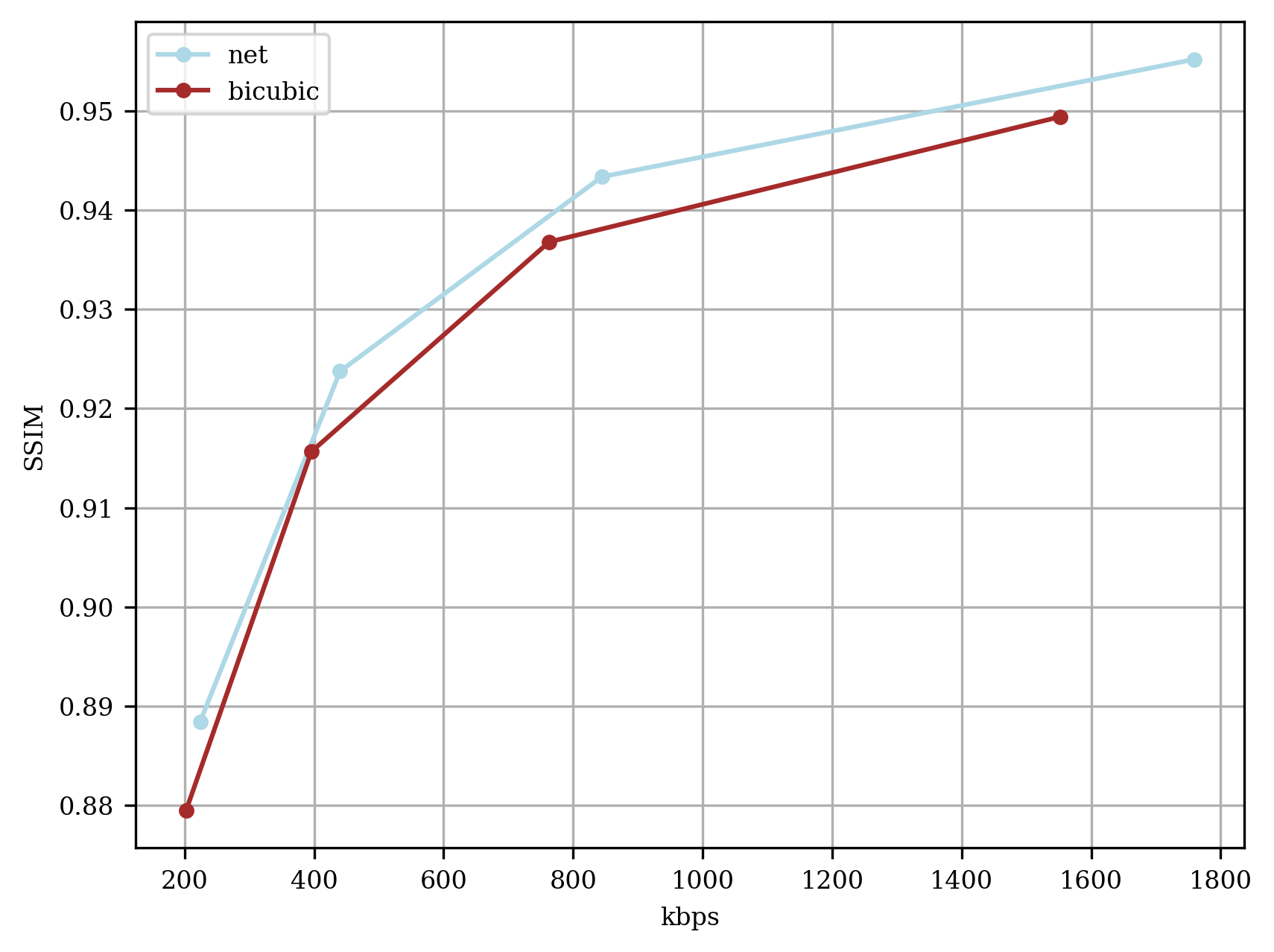}
\label{xiph-ssim-266}
\end{minipage}%
}
\subfigure[BD-PSNR of XIPH]{
\begin{minipage}[t]{0.4\linewidth}
\centering
\includegraphics[width=1.3in]{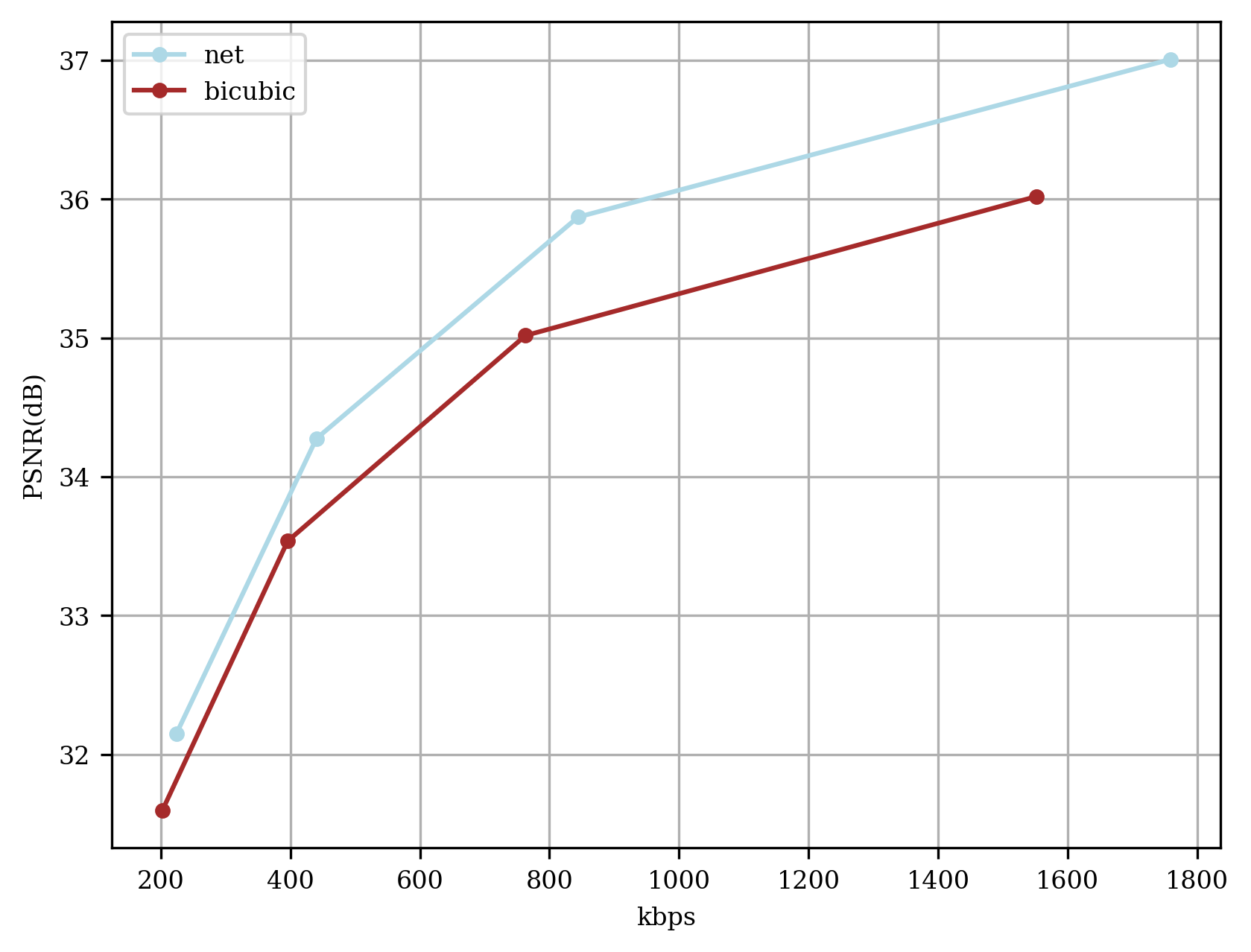}
\label{xiph-psnr-266}
\end{minipage}%
}
\caption{\centering  Comparison of proposed method and liner filter baseline on XIPH dataset of H.266/VVC codec.}
\label{bdcurve-266}
\end{figure}

\begin{figure}[htbp]
    \centering
    \subfigure[sergey-zolkin]{
        \includegraphics[width=0.4\linewidth]{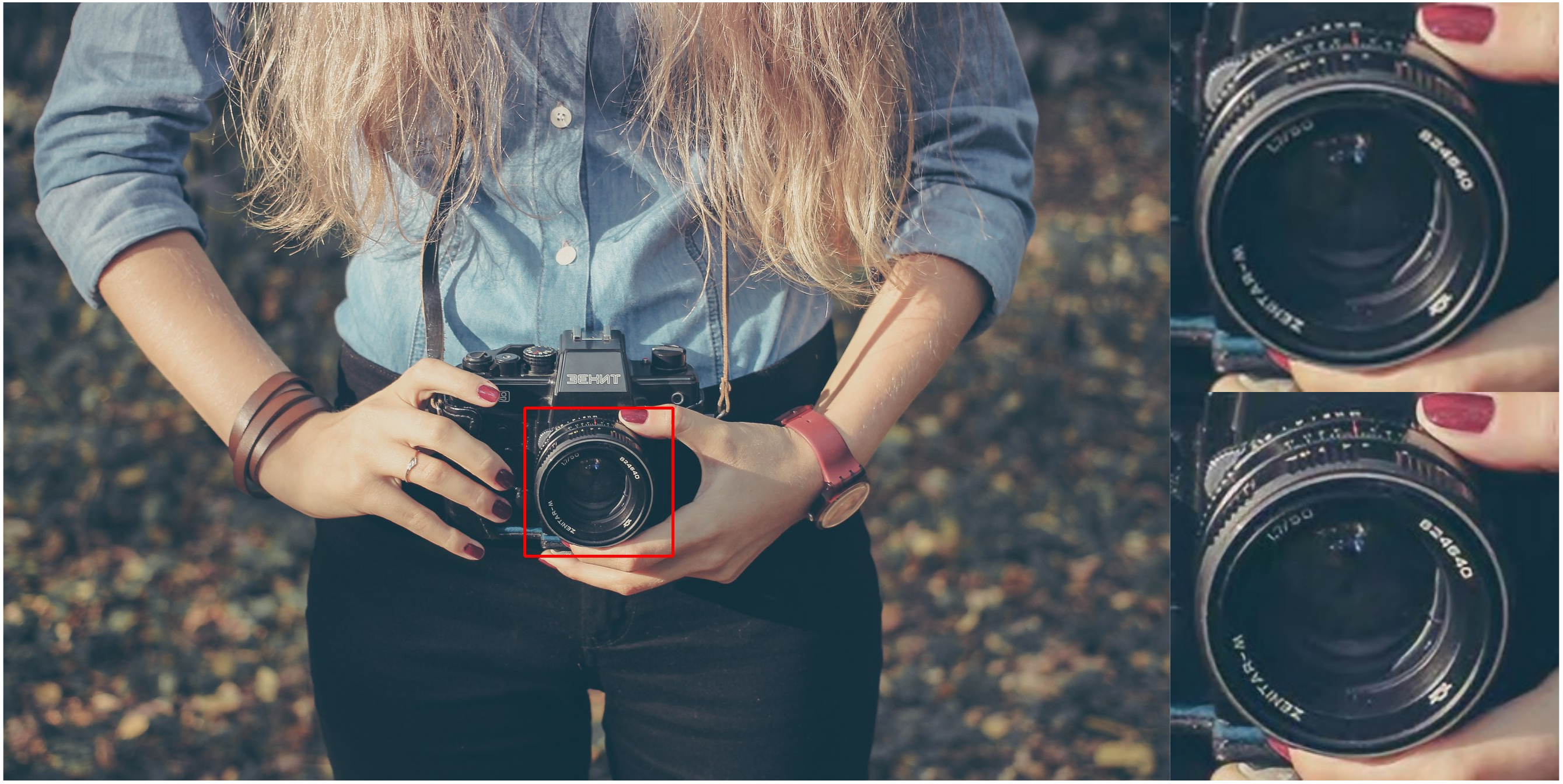}
        \label{sergey-zolkin}
    }
    \subfigure[zuger]{
	\includegraphics[width=0.4\linewidth]{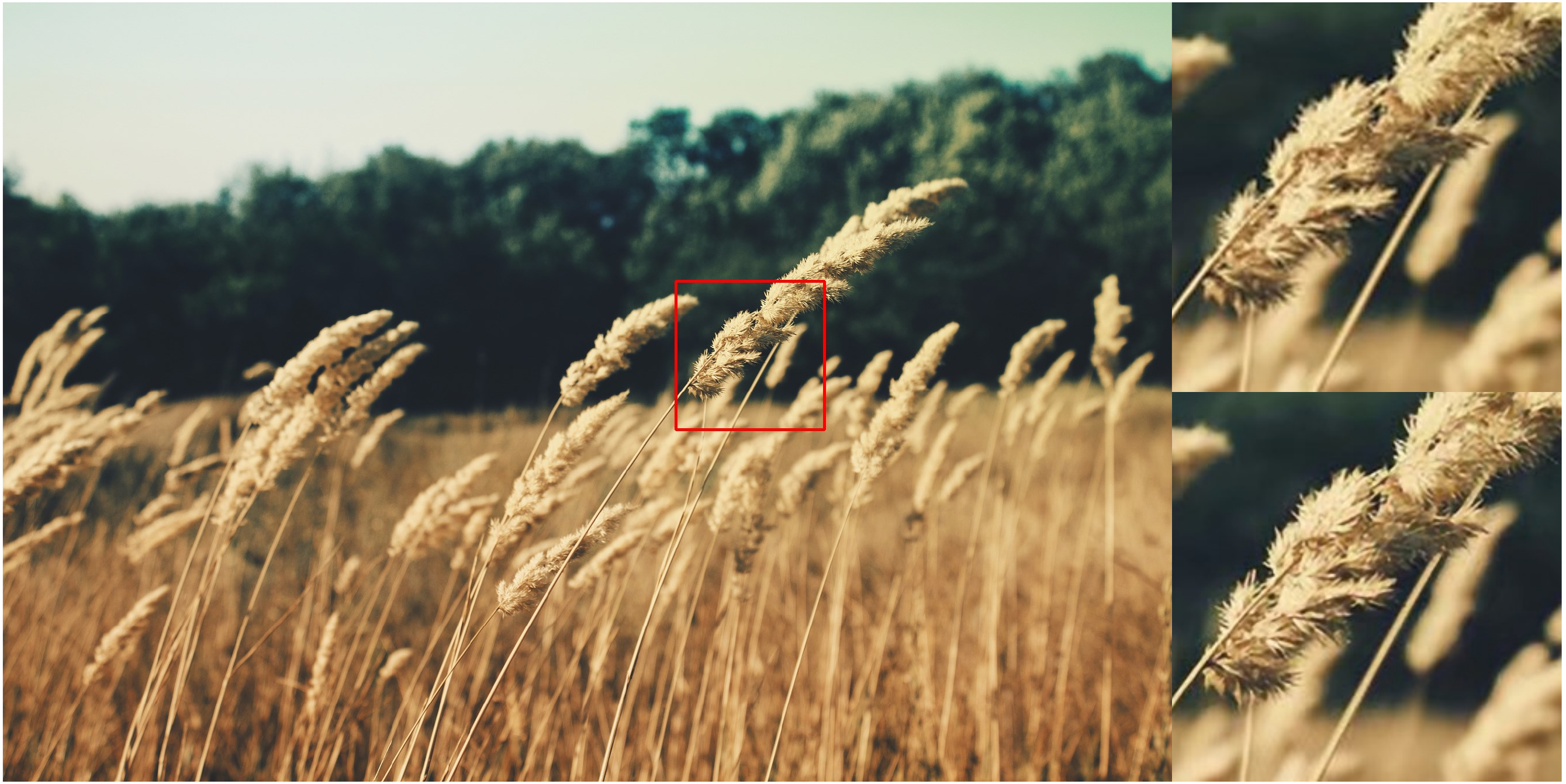}
        \label{zuger}
    }
    \quad
    \subfigure[vita-vilcina]{
    	\includegraphics[width=0.4\linewidth]{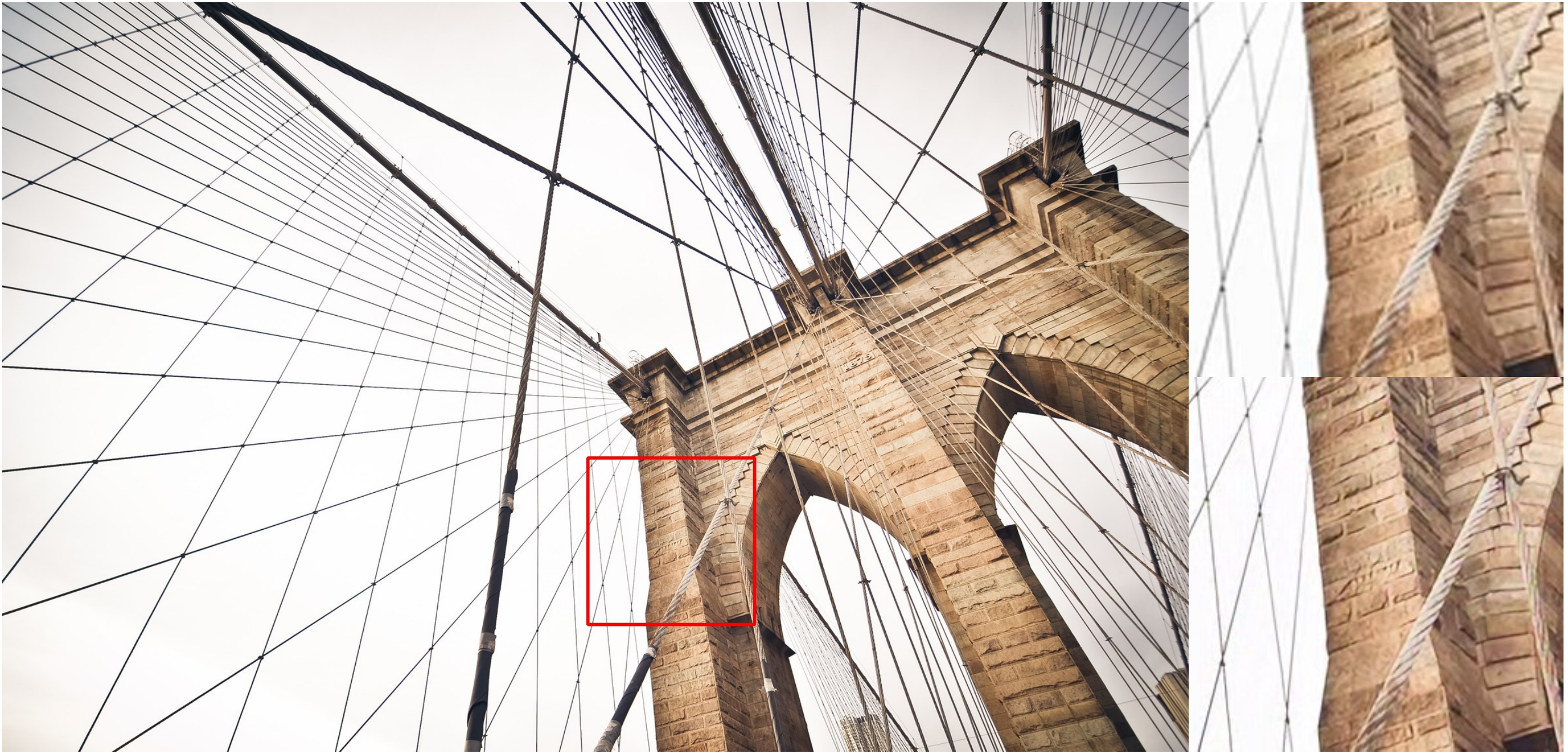}
        \label{vita-vilcina-bic}
    }
    \subfigure[thong-vo]{
	\includegraphics[width=0.4\linewidth]{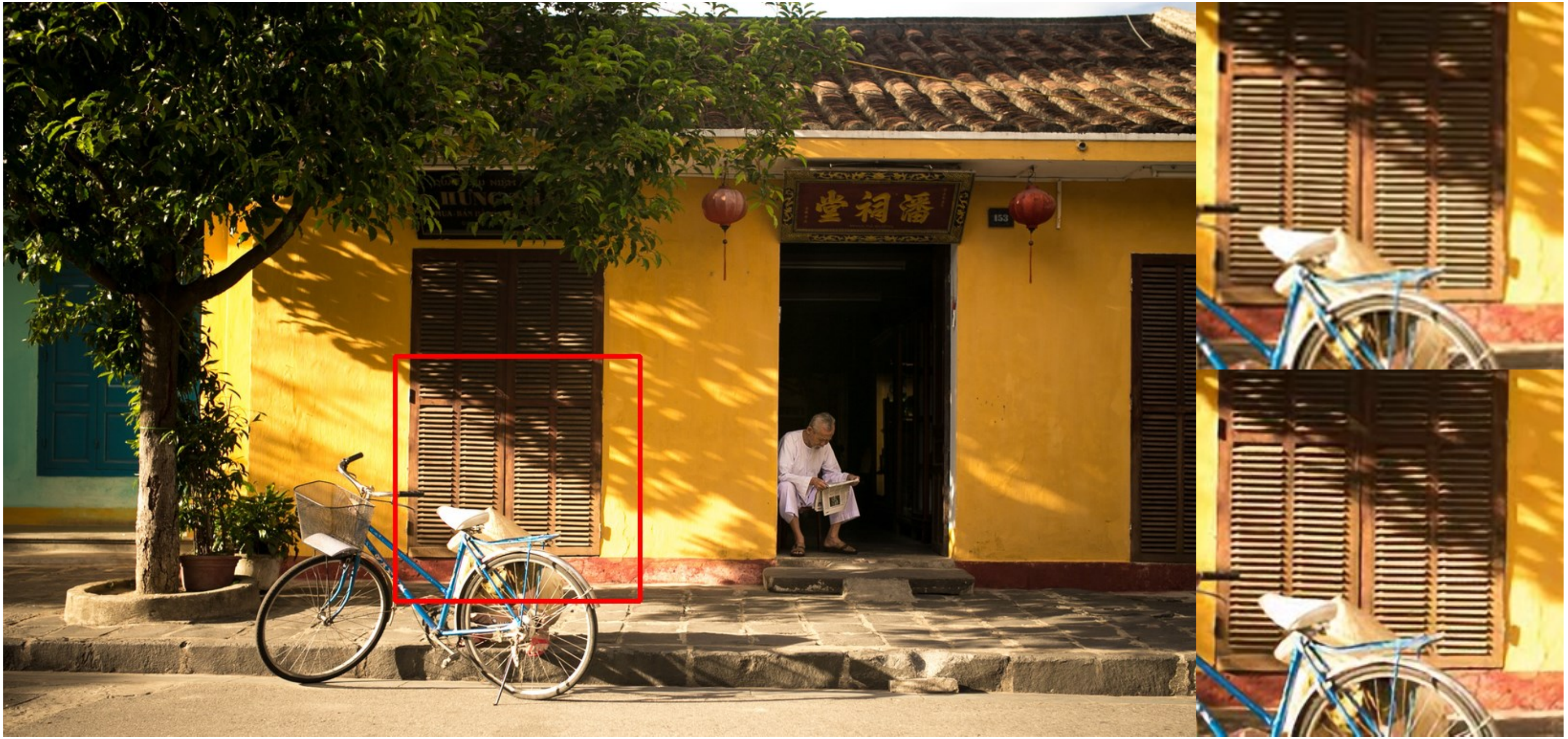}
        \label{thong-vo}
    }
    \caption{\centering  Qualitative results of our models on CLIC compression datasets. The top right is Bicubic, and the bottom right is RARN in terms of downsampling methods.}
    \label{Qualitative_results}
\end{figure}

\begin{table}[t]
\begin{center}
\caption{\centering Average BD-Rate  ($\Delta \mathrm{R}$), BD-PSNR ($\Delta \mathrm{P}$), BD-SSIM ($\Delta \mathrm{S}$) results on XIPH dataset of H.266/VVC. }
\label{vvc}
{
\begin{tabular}{ccccc}
\hline 
 encoding configuration & $\Delta \mathrm{R}$ & $\Delta \mathrm{P}$ & $\Delta \mathrm{R}$ & $\Delta \mathrm{S}$ \\
\hline 
{\makecell[c]{CQP, preset=faster}} 
& $-13.98\%$ & $0.36\mathrm{~dB}$ & $-6.32\%$ & $4e-3$ \\
\hline
\end{tabular}
}
\end{center}
\end{table}

\subsection{Dataset}
We use the large-scale high-quality video dataset Vimeo90K \cite{xue2019video} dataset for training.
In order to be consistent with previous approaches, we use two publicly available datasets for validation.
For precoding, we used the videos in 1080P and yuv420p format from the XIPH \cite{montgomery1994xiph} dataset.
For the task of jointly upsampling and downsampling, we use the HEVC standard test sequences \cite{sullivan2012overview} with five categories of videos.

\subsection{Results of Precoding and Qualitative Results}
We use the proposed RARN for precoding original videos into low-resolution frames, using the single model to convert the video to different resolutions.
The low-resolution video is restored to its original resolution using the linear filter supported in the codec after encoding and decoding.
We use 1080p FHD yuv420p video in XIPH for testing, and the benchmark utilizes Bicubic/Lanczos sampling as precoding.
The standard codec is $libx265$ in $FFmpeg$, and we use the "medium" preset, two pass rate control modes, GOP = 30, and a bitrate range of $0.5\overline{\ }10$Mbps.
We calculated the Bjøntegaard Delta Bit-Rate (BDBR) in the y channel with PSNR as the quality metrics.
The bitrate savings at different rescaling factors are shown in Table~\ref{xiph_265}, and we achieve better results than previous methods in most scenarios.
Specifically, the RD curves of video old\_town\_cross in the XIPH dataset with Lanczos downsampling as the baseline under $\times2$ scale factor are shown in Fig.~\ref{xiph-ssim} and Fig.~\ref{xiph-psnr}.

Furthermore,  we implement experiments on H.266/VVC codec to demonstrate that our proposed algorithm framework is compatible with different codecs.
Quantitative results in Table~\ref{vvc} are anchored by $\times2$ Bicubic downsampling, using the faster preset CQP encoding configuration. The QP of the VVC codec is set to 22, 32, 37, and 42. The RD curves of video $controlled\_burn$ are shown in Fig.~\ref{xiph-ssim-266} and Fig.~\ref{xiph-psnr-266}.

To characterize the visual quality improvement of the proposed algorithm, we use the validation images of the CLIC dataset for visualization.
We perform the down-sampling image compression of $\times2$ magnification in the HEVC All Intra (AI) configuration with $QP$ set to 22 and Bicubic upsampling.
Precoding with RARN enables us to improve the quality of the restored images from 35.1dB to 35.6dB for PSNR and from 9.89 to 9.90 for MS-SSIM on average.
The reconstructed images for qualitative comparison are shown in Fig.~\ref{Qualitative_results}. 

\subsection{Results of Joint Processing}
In downsampling-based video compression schemes, super-resolution networks can be used as post-processing by jointly training with virtual codec networks \cite{son2021enhanced, wei2021video}.
We use a jointly optimized symmetric RARN as post-processing for a fair comparison.
We test on the standard test sequence of HEVC and calculate the BD-rate under the HEVC ALL Intra (AI) encoding configuration.
Experimental results in Table~\ref{HEVC} demonstrate that we achieve better compression performance on HEVC class B, C, D, and E.
The compression performance metrics measured by BD-SSIM and BD-PSNR of the video basketball\_drive of ClassB are shown in Fig.~\ref{hevc-ssim} and Fig.~\ref{hevc-psnr}, which is anchored by Lanczos $\times2$ downsampling in y channel.

\subsection{Ablation Study and Inference Speed}

\begin{table}[t] \small
\begin{center}
\caption{\centering Ablation Study on XIPH Dataset}
\label{ablation}
{
\begin{tabular}{|c|c|}
\hline scale factor $\times2$ & BD-Rate (Lanczos) \\
\hline proposed & $\mathbf{-16.24 \%}$ \\
\hline RARN wo/ sampling compensation & $-14.55 \%$ \\
\hline RARN wo/ auxiliary bitrate information & $-9.35 \%$ \\
\hline TVC wo/ cyclic shift & $-13.94 \%$ \\
\hline TVC wo/ INN structure & $-12.48 \%$ \\
\hline
\end{tabular}
}
\end{center}
\end{table}

To demonstrate the effectiveness of key components of the proposed model, we conduct experiments in Table~\ref{ablation}.
We use the FHD videos in the XIPH dataset and employ fixed Lanczos $\times2$ downscaling as a benchmark.
We calculate BD-rate with the $medium$ preset, two pass rate control modes, GOP = 30, and bitrate range of $0.5\overline{\ }10$Mbps.
We remove the bitrate auxiliary information and the sampling compensation MLP in the precoding module, and both resulted in performance degradation.
The bitrate prior probability is revealed to contribute to a significant performance gain. The cyclic shift structure in TVC can also enhance the network's ability to simulate the codec's behavior and bring performance gains.
In addition, we replace the INN structure with a simple CNN resulting in a huge performance decline, indicating that the proposed structure performs well when used for codec simulation.

By removing the double-headed MLP in the sampling compensation module to lighten the model structure, we reduce the parameter size of RARN from $9.3M$ to $3.4M$ and still maintain a relatively competitive performance (RARN wo/ sampling compensation in the ablation study). The processing speed for 1080p video on a single NVIDIA V100 GPU has also increased from 20.1fps to 90.9fps, satisfying the real-time inference requirements.

\section{Conclusion}
We propose a rate-guided arbitrary rescaling network (RARN) to preprocess videos to improve compression performance and achieve arbitrary resolution conversion with a single model.
The method is fully compatible with standard codecs and transmission pipelines and is optimized end-to-end based on RD loss. 
We also propose an iteratively optimized transformer-based virtual codec (TVC) to simulate the degradation process of the standard codec and provide an accurate gradient estimation.
The experimental results demonstrate that we achieve better compression performance over previous methods under various rescaling scales and encoding configurations and can achieve real-time precoding with satisfactory performance. In the following research, we will further input the upsampling method signal into RARN as a condition for using a single RARN to adapt to different upsampling methods.

\Section{References}
\bibliographystyle{IEEEbib}
\bibliography{refs}

\end{document}